\documentclass[aip,reprint]{revtex4-1}
\usepackage{graphicx}
\usepackage{dcolumn}
\usepackage{bm}
\usepackage{amsmath}

\draft 

\begin{document}

\title{Heat transport of nitrogen in helium atmospheric pressure microplasma} 



\author{S. F. Xu}
\noaffiliation
\affiliation{Key Laboratory for Laser Plasmas (Ministry of Education) and Department of Physics and Astronomy, The State Key Laboratory on Fiber Optic Local Area, Communication Networks and Advanced Optical Communication Systems, Shanghai Jiao Tong University, Shanghai 200240, China}
\author{X. X. Zhong}
\email{xxzhong@sjtu.edu.cn}
\noaffiliation
\affiliation{Key Laboratory for Laser Plasmas (Ministry of Education) and Department of Physics and Astronomy, The State Key Laboratory on Fiber Optic Local Area, Communication Networks and Advanced Optical Communication Systems, Shanghai Jiao Tong University, Shanghai 200240, China}


\date{\today}

\begin{abstract}
Stable DC atmospheric pressure normal glow discharges in ambient air were produced between the water surface and the metallic capillary coupled with influx of helium gas. Multiple independent repeated trials indicated that vibrational temperature of nitrogen rises from 3200 to 4622 K, and rotational temperature of nitrogen decreases from 1270 to 570 K as gas flux increasing from 20 to 80 sccm and discharge current decreasing from 11 to 3 mA. Furthermore, it was found that the vibrational degree of the nitrogen molecule has priority to gain energy than the rotational degree of nitrogen molecule in nonequilibrium helium microplasma.
\end{abstract}


\maketitle 

Microplasma has received considerable attention in recent years, due to its numerous potential applications in biomedicine, nanomaterial processing and environmental protection.\cite{Moselhy2002,Iza2003,Sankaran2003,Kikuchi2004,Ostrikov2005,Richmonds2008,Levchenko2008,Xiong2010,Mariotti2010,Mariotti2011,Ostrikov2013} Manifold devices are used to generate stable microdischarges, inluding DC, RF, Microwave, Laser etc.\cite{Barbeau1990,Baba2007,Kim2005,Nedanovska2011} based on the different applied operation modes. Among all kinds of non-equilibrium plasma systems, the microplasma in close proximity to the water surface is promising in fabrication and engineering of nanoparticles and quantum dots as shown in the previous research.\cite{Mariotti2010,Mariotti2011,Miron2011,Naoki2011,Li2012,Huang2013}

Microplasma characteristics such as gas temperature, electron temperature, electron density, electric field etc. play an important role in physical and chemical process. In particular, gas temperature, the basic thermodynamic state parameter, decides the neutral gas density with state equation and helps us understand the power consumption.\cite{Donnelly2000,Hash2001} Gas temperature is always estimated using radiative transitions of diatomic molecules, such as $N_{2},\ C_{2},\ OH,\ CO$, whose rotational temperature was considered to be equal to gas translational temperature.\cite{Faure1998,Cruden2002,Staack2005,Bai2006,Raud2011} It is worthwhile to note that the ratio of vibrational temperature to rotational temperature can suggest nonequilibrium level of microplasma.\cite{Staack2005}

This paper reports the rotational and vibrational tempeture of nitrogen under the controlled conditions, discharge current (from 2 to 11 mA) and gas flux (from 20 to 80 sccm) on assumpation that rotational and vibrational states obey Boltzmann distribution. Further more, we calculate and compare the rotational and vibrational inner energy increment under the same assumpation.

\begin{figure}
\center
\includegraphics{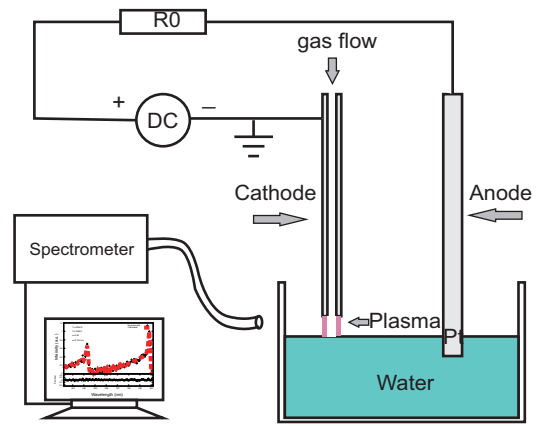}
\caption{Schematic diagram of the experimental set-up.}
\label{device}
\center
\end{figure}

A schematic diagram of the experimental set-up is shown in Fig.~\ref{device}. The circuit is driven by a high-voltage dc source (purchased from Dongwen Corporation in Tianjin, China), with a platinum electrode immersed in water acting as the anode, and a stainless-steel capillary of 175 $\mu$m internal diameter working as cathode. A ballast resistor (about 60 k$\Omega$) limits the discharge current and provides stability to the discharge. Helium flow through a mass flow controller is introduced into the capillary as the working gas. The gap between the capillary and the surface of the water is adjusted to 1 mm. The plasma is ignited when the voltage is increased to about 2000 V. Thereafter the voltage decreases to about 600 V and reaches stable conditions at currents of milliamperes. The two controlling parameter discharge current $I$ varies from 2 to 11 mA and gas flux $flux$ varies from 20 to 80 sccm. The mean speed of gas flow nearly equals $0.693\times flux$ m/s ($flux$ in sccm), varies from 13.86 to 55.44 m/s. Reynolds coefficient varies from 20 to 92, so the gas motion is stratified flow rather than turbulent flow. Emission spectroscopy irradiated by excited nitrogen were collected by the AvaSpec-2048FT-4-DT spectrometer.

The vibrational and rotational temperatures are estimated comparing experimental spectrum and calculated spectrum of second positive systems of $N_{2}$ assuming Boltzmann distribution of vibrational and rotational states population. The temperatures assuring the best fitting between experimental and model spectra are taken as vibrational and rotational temperatures\cite{Wang2005,Poirier2011,Raud2011}.

We just consider $N_2(C^3\Pi_u,v'=2\to B^3\Pi_g,v''=5)$ and $N_2(C^3\Pi_u,v'=1\to B^3\Pi_g,v''=4)$ emissions from 392 nm to 400 nm where $C^3\Pi_u,\ B^3\Pi_g$ denote electron configuration, $v',\ v''$ denote vibrational states. Rotational and vibrational bands come from the transitions of different rotational and vibrational states($v'J'\to v''J''$). Emission intensity is proportional to spontaneous emission coefficient and population of up state,
\begin{equation}
I_{Bv''J''}^{Cv'J'}=N_{v'J'}\frac{hc}{\lambda_{v'J'v''J''}}A_{v'J'v''J''},
\end{equation}
where $A_{v'J'v''J''}$ is spontaneous emission coefficient, $N_{v'J'}$ is population of up state, $\lambda_{v'J'v''J''}$ is wavelength, $v'J'$ denote up vibrational and rotational state, $v''J''$ denote low vibrational and rotational state, $h$ is the planck constant, $c$ is the speed of light.

Under the Born-Oppenheimer approximation and Boltzmann distribution assumption, the intensity can be written as 
\begin{equation}
I_{Bv''J''}^{Cv'J'}=\frac{D}{\lambda^{4}}q_{v',v''}exp(-\frac{E_{v'}}{kT_{v}})S_{J',J''}exp(-\frac{E_{J'}}{kT_{r}}),
\label{In}
\end{equation}
where $D$ is constant, $q_{v',v''}$ is the Franck-Condon factor given by Hartmann,\cite{Hartmann1978} the line strength intensity $S_{J',J''}$ is given by Phillips,\cite{Phillips1976} vibrational state energy $E_{v'}$ and rotational state energy $E_{J'}$ are calculated as did by Bai.\cite{Bai2006}

\begin{figure}
\center
\includegraphics{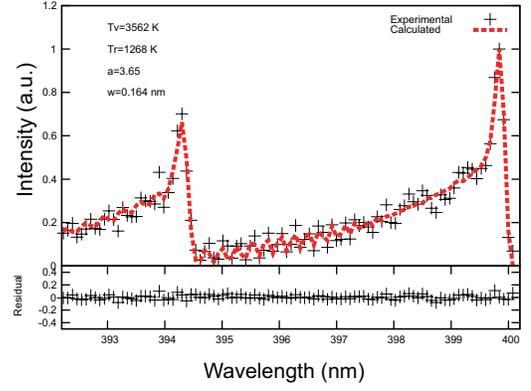}
\caption{Theoretical fitting of measured rovibrational bands of $N_{2}(C^{3}\Pi_{u}\to B^{3}\Pi_{g})$ second positive system, which is in the wavelength range of 392 to 400 nm. The discharge current is $10\ mA$, gas flux is $40\ sccm$.}
\label{fit}
\center
\end{figure}

The procedure for obtaining rotational and vibrational temperatures involves 4 steps. Firstly, we calculate emission wavelength by using Eqs.~\ref{WL}.
\begin{equation}
\begin{split}
\lambda^{Cv'J'}_{Bv''J''}=\{n_a\sum_{p=0}^5\sum_{q=0}^2Y^C_{pq}(v'+\frac{1}{2})^p[J'(J'+1)]^q \\
-Y^B_{pq}(v''+\frac{1}{2})^p[J''(J''+1)]^q\}^{-1}\\
\end{split}
\label{WL}
\end{equation}
where $n_a$ is the index of refactivity, $Y_{pq}$ are constants given by Bai\cite{Bai2006}. Secondly, we calculate the line indensity using Eqs.~\ref{In}. Thirdly, emission lines are convoluted with point-spread function $g(\Delta\lambda)$ considering instrumental effect, Dopler broadening and Stark broadening effect.\cite{Phillips1976}
\begin{equation}
g(\Delta\lambda)=\frac{a-(2\Delta\lambda/w)^2}{a+(a-2)(2\Delta\lambda/w)^2},
\end{equation}
It has a maximum $g(0)=1$, a full width half maximum $w$, and the wings extend to $\pm wa^{1/2}$. At last, we use Newton-Raphson method to optimize the fitting of calculated spectrum and experimental spectrum, and get $T_{v},\ T_{r}$ broadening parameters a, w at the same time. Fig.~\ref{fit}. shows an example of the $N_{2}$ temperature measurement by fitting one rovibrational bands, discharge current $I=10$ mA, $flux=40$ sccm, and the optimum results are $T_{v}=3562$ K, $T_{r}=1268$ K.

\begin{figure}
\center
\includegraphics{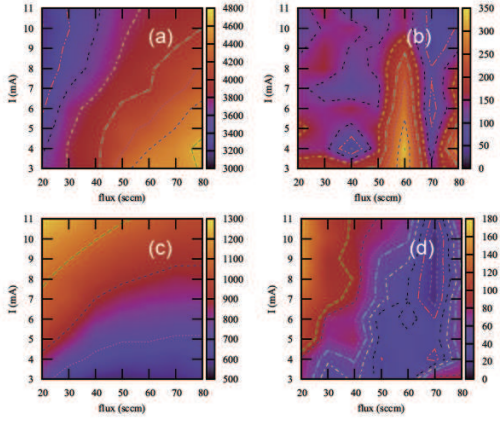}
\caption{Vibrational and rotational tempeture (in K) maps versus discharge current and gas flux. (a) statistical mean of $T_{v}$, (b) statistical standard deviation of $T_{v}$, (c) statistical mean of $T_{r}$, (d) statistical standard deviation of $T_{r}$.}
\label{mfiT}
\center
\end{figure}

\begin{figure}
\center
\includegraphics{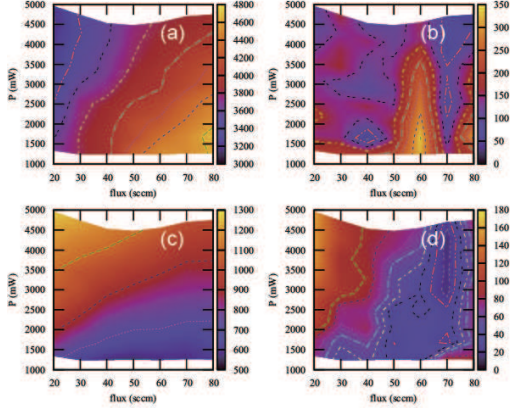}
\caption{Vibrational and rotational tempeture (in K) map versus mean Power and gas flux. (a) statistical mean of $T_{v}$, (b) statistical standard deviation of $T_{v}$, (c) statistical mean of $T_{r}$, (d) statistical standard deviation of $T_{r}$.}
\label{mfpT}
\center
\end{figure}

We use the two-way ANOVA (analysis of variance) to check significance about factors' influence on temperautres. Table~\ref{ttv}, ~\ref{ttr} show the results of ANOVA, where gas flux is the main factor B and discharge current is the main factor A. Both gas flux and discharge current significantly affect vibrational temperature and rotational temperature, but their mixed effect is not significant. Vibrational temperature is positively correlated with gas flux and negatively correlated with discharge current (power), on the contrary, rotational temperature is negatively correlated with gas flux and positively correlated with discharge current (power) as shown in FIG.~\ref{mfiT}, ~\ref{mfpT}. Multiple independent repeated trials (six times change $I$ from 2 to 11 mA, $flux$ from 20 to 80 sccm) show that vibrational temperature rises from 3200 to 4622 K, in contrast, rotational temperature decreases from 1270 to 570 K with decreasing discharge current and incresing gas flux. The ratio of vibrational and rotational temperature rises from 2.6 to 7.8 with decreasing discharge current and incresing gas flux. The microplasma becomes more non-equilibrium by decreasing discharge current and increasing gas flux.
\begin{table}
\caption{Two-way ANOVA of $T_v$}
\label{ttv}
\begin{ruledtabular}
\begin{tabular}{ccccc}
& SumOfSq & DF & MeanSq & FRatio\\
\hline
$A(I)$ & $1.47\times 10^7$ & 8 & $1.83\times 10^6$ &66.51\\
$B(flux)$ & $3.73\times 10^7$ & 6 & $6.22\times 10^6$ & 225.72\\
$A\times B$ & $1.12\times 10^6$ & 48 & $2.33\times 10^4$ & 0.85\\
Eeror & $8.68\times 10^6$ & 315 & $2.76\times 10^4$ &\\
Total & $6.18\times 10^7$ & 377 & &\\
\end{tabular}
\end{ruledtabular}
\end{table}
\begin{table}
\caption{Two-way ANOVA of $T_r$}
\label{ttr}
\begin{ruledtabular}
\begin{tabular}{ccccc}
 & SumOfSq & DF & MeanSq & FRatio\\
\hline
$A(I)$ & $1.01\times 10^7$ & 8 & $1.27\times 10^6$ &208.31\\
$B(flux)$ & $2.74\times 10^6$ & 6 & $4.56\times 10^5$ & 75.00\\
$A\times B$ & $2.17\times 10^5$ & 48 & $4.53\times 10^3$ & 0.74\\
Eeror & $1.92\times 10^6$ & 315 & $6.08\times 10^3$ &\\
Total & 1.50$\times 10^7$ & 377 & &\\
\end{tabular}
\end{ruledtabular}
\end{table}

Diatomic molecule energy includes translational energy, vibrational energy and  rotational energy
\begin{equation}
\varepsilon=\varepsilon^{t}+\varepsilon^{v}+\varepsilon^{r}.
\end{equation}
Characteristic vibrational and rotational temperature of $N_{2}$ are $\theta_{v}=3340 \ K,\theta_{r}=2.86 \ K$(far less than rotational temperature), so effect of homonuclear molecule microparticles identity on rotational states is neglected. Under Boltzman distribution assumption, vibrational and rotational energy, specific heat capacity at constant volume are 
\begin{eqnarray}
& & U^{v}=Nk(\frac{\theta_{v}}{2}+\frac{\theta_{v}}{e^{\frac{\theta_{v}}{T_{v}}}-1}), C_{v}^{v}=Nk(\frac{\theta_{v}}{T_{v}})^{2}\frac{e^{\frac{\theta_{v}}{T_{v}}}}{(e^{\frac{\theta_{v}}{T_{v}}}-1)^{2}}, \\
& & U^{r}=NkT_{r}, C_{v}^{r}=Nk,
\end{eqnarray}
where N is molecule number, k is Boltzmann constant.
Because the vibrational capacity is far less than rotational capacity, vibrational energy increment is far less than rotational energy increment in a reversible thermodynamic process as temperature going up. On our experimental conditions, final state is not a complete thermodynamic equilibrium state where translational temperature, vibrational temperature and rotational temperature are not all equal. Main excited processes of nitrogen are the inelastic collisions between $N_2$ and He, electron as shown by Wang.\cite{Wang2005}

\begin{figure}
\center
\includegraphics{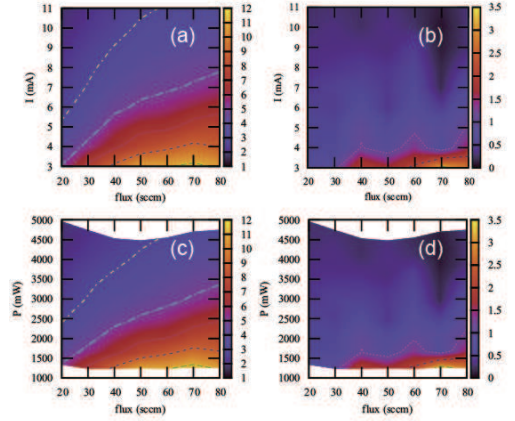}
\caption{Energy ratio of vibrational energy increment to rotational energy increment. (a) statistical mean of energy ratio versus discharge current and gas flux, (b) statistical standard deviation of energy ratio versus discharge current and gas flux, (c) statistical mean of energy ratio versus mean Power and gas flux, (d) statistical standard deviation of energy ratio versus mean Power and gas flux.}
\label{mie}
\center
\end{figure}

Temperature is a thermodynamic equilibrium concept and combined with innernal energy. Bolzmann distribution assumption was used to get vibration and rotational temperatures, so it is reasonable that rotational and vibrational internal energy increment can be calculated integrating capacity through temperature path respectively. The molecule number of $N_{2}$ is an unknown factor which decides absolute internal energy, but it makes no difference to the ratio of vibrational energy increment to rotational energy increment (called energy ratio). The energy ratio is large than 1.0 and goes up from 1.9 to 12 with decreasing current and incresing gas flux as shown in Fig.~\ref{mie}. It can be noticed that the vibrational degree has priority to get energy over the rotational degree when plasma is nonequilibrium. Particle simulation will be needed to check this directly.


In a wide range parameter space (discharge current varies from 2 to 11 mA, gas flux varies from 20 to 80 sccm), stable microdischarge is maintained, and nitrogen vibrational tempeture, rotational temperature and energy ratio maps in dependence of the gas flux and discharge current are obtained. Multiple independent repeated trials show that vibrational temperature rises from 3200 to 4622 K, in contrast, rotational temperature decrease from 1270 to 570 K, the energy ratio of vibrational energy increment to rotational energy increment goes up from 1.9 to 12 with decreasing discharge current (11 to 3 mA) and incresing gas flux (20 to 80 sccm). The microplasma gets more nonequilibrium when discharge current becomes smaller and gas flux becomes stronger, and the degree of vibration has priority to get energy over the degree of rotation in nonequilibrium helium system. The result sheds light on the mechanism of the heat transport between the vibrational degree and the rational degree of the nitrogen molecule in the nonequilibrium helium gas microplasma.

\begin{acknowledgments}
XXZ and SFX acknowledge support from the NSFC (Grant No. 11275127, 90923005), STCSM (Grant No. 09ZR1414600) and MOST of China. 
\end{acknowledgments}

\providecommand{\noopsort}[1]{}\providecommand{\singleletter}[1]{#1}%

\end{document}